\documentstyle[nato,epsf]{crckapb}
\begin{opening}
\title{Some thermodynamic aspects of self-assembly of arrays of quantum dots}
\author{Jos\'e Emilio Prieto}
\institute{Institut f\"ur Experimentalphysik, Freie Universit\"at Berlin \\
Arnimallee 14, 
14195 Berlin, Germany}
\author{Ivan Markov}   
\institute{Institute of Physical Chemistry,\\ Bulgarian Academy of Sciences\\ 
1113 Sofia, Bulgaria}
\end{opening}
\begin{document}
\begin{abstract}
We have studied the relative adhesion (the wetting) of disloca\-tion-free
three-dimensional islands belonging to an array of islands to the wetting 
layer in Stranski-Krastanov growth mode. 
The array has been simulated as a chain of islands in 1+1
dimensions placed on top of a wetting layer. In addition to the critical size
of the two-dimensional islands for the 2D-3D transformation to occur, we 
find that the wetting depends strongly on the density of the array, the
size distribution and the shape of the islands.

\end{abstract}

\section{Introduction}

The instability of planar films against coherently strained three-dimensional 
(3D) islands in highly mis\-matched epitaxy is a subject of intense research 
in recent time owing to their possible optoelectronic
applications as quan\-tum dots.\cite{Politi} The term ``coherent
Stran\-ski-Krastanov (SK) growth'' has been coined for this case of formation
of 3D islands that are strained to fit the underlying wetting layer at the
interface but are largely strain-free near their top and side
walls.\cite{Eag,Vit} This term was introduced in order to distinguish this
case from the ``classical'' SK growth in which the lattice misfit is
accommodated by misfit dislocations at the interface.\cite{Matt}

Experimental studies of arrays of coherent 3D islands in SK growth of highly
mismatched se\-mi\-con\-duc\-tor ma\-te\-ri\-als one on top of the other have
shown surprisingly narrow size distributions of the
is\-lands,\cite{Leo,Moi,Gru,Jia} (see also Ref. \cite{Vit} and the
references therein). (It is worth noting that a narrow size distribution
has been established also in the Volmer-Weber growth of metals on insulators
in the absence of a wetting layer.\cite{zz}) This phenomenon, known in
the literature as self-assembly (for a review see Ref. \cite{Chris}), is
highly desirable as it guarantees a spe\-ci\-fic optical wavelength of the
array of quantum dots. The physics of this self-assembly is still not
understood in spite of the numerous thermodynamic and kinetic
studies.\cite{ter,sch,dobbs,zang} For example, Priester and Lannoo found that
two-dimensional (2D) islands with a monolayer height act as precursors of
the 3D pyramidal islands,\cite{Pri} (see also Ref. \cite{chen}). The energy
per atom of the 2D islands possesses a minimum for a certain volume, but
the 3D islands become energetically favored at a smaller size.
Thus, at some critical surface coverage, the 2D islands spontaneously transform
into 3D islands preserving a nearly constant volume during the 2D-3D
transformation. The resulting size distribution reflects that of the 2D
islands which is very narrow. This picture has been recently cor\-ro\-bo\-rated
by Ebiko {\it et al.}\cite{Ebi} who found that the volume distribution of
InAs/GaAs self-assembled quantum dots agrees well with the scaling function
that is characteristic for the two-dimensional submonolayer homoepitaxy.
\cite{Amar} Korutcheva {\it et al.}\cite{Kor} and Markov and
Prieto \cite{Marjosem} reached the same conclusion with the exception that the
2D-3D transformation was found to take place through a series of
in\-ter\-mediate states with discretely increasing thickness (one, two,
three, etc. monolayers-thick islands) that are stable in separate consecutive
intervals of volume. Khor and Das Sarma arrived to the same conclusion by
using Monte Carlo simulations.\cite{Khor}

In a recent paper, Prieto and Markov discussed the formation of coherent 3D
islands within the framework of the traditional concept of
wetting.\cite{Prieto} As is well known the wetting parameter which accounts
for the energetic influence of a crystal B in the heteroepitaxial growth 
of a crystal A on top of it is defined as (for a review see Ref. \cite{Mar2})
\begin{equation}
\Phi  = 1 - \frac{E_{\rm AB}}{E_{\rm AA}}
\end{equation}
where $E_{\rm AA}$ and $E_{\rm AB}$ are the energies per atom required to
disjoin a half-crystal A from a like half-crystal A and from an unlike
half-crystal B, respectively.

The mode of growth of a thin film is determined by the difference $\Delta \mu
= \mu (n) - \mu _{\rm 3D}^0$, where $\mu (n)$ and $\mu _{\rm 3D}^0$ are the
chemical potentials of the film (as a function of its thickness $n$) and of the
bulk material A, respectively.\cite{Mar2} 
The chemical potential of the bulk crystal A is
given at zero temperature by the work $\phi _{\rm AA}$ (taken with a negative
sign) to detach an atom from the well known kink or half-crystal position. 
The latter name is due to the fact that an atom at this position is bound 
to a half-atomic row, a half-crystal plane and
a half-crystal block.\cite{Koss,Stran}
In the case of a monolayer-thick film of A on the surface of B the chemical
potential of A is given by the analogous work $\phi _{\rm AB}$ with the
exception that the underlying half-crystal block of A is replaced by a
half-crystal block of B. Thus $\Delta \mu  = \phi _{\rm AA}
- \phi _{\rm AB}$. In the simplest case of additivity of bond energies the
difference $\phi _{\rm AA} - \phi _{\rm AB}$ reduces to $E_{\rm AA} -
E_{\rm AB}$ as the lateral bondings cancel each other. Then $\Delta \mu $ is
proportional to $\Phi $, i.e. $\Delta \mu  = E_{\rm AA}\Phi $.\cite{Mar2} It 
follows that it is the wetting parameter $\Phi $ which 
determines the mechanism of growth of A on B.\cite{Rudy} In the two limiting
cases of growth of isolated 3D islands of A directly on the surface of B
[Volmer-Weber growth, characterized by incomplete wetting ($0 < \Phi  < 1$) 
and any misfit $\varepsilon _0 = \Delta a/a$], or by consecutive formation of
monolayers of A on B [Frank - van der Merwe growth, with complete wetting
($\Phi \le 0$) and $\varepsilon _0 \approx 0$], $\Delta \mu $ goes
asymptotically to zero from above and from below, respectively, but it changes
sign in the case of growth of 3D islands of A on a thin wetting layer of A on
the substrate B [Stranski-Krastanov growth, complete wetting at the
beginning ($\Phi  \le 0$) and $\varepsilon _0 \ne 0$].\cite{Prieto,Mar2} 
Equation $\Delta \mu  =
E_{\rm AA}\Phi $ is thus equivalent to the familiar 3-$\sigma $ criterion of
Bauer.\cite{Mar2,Bau}

The Stranski-Krastanov morphology appears as a results of the interplay of the
film-substrate bonding, misfit strain and the surface energies. A wetting layer
with a thickness of the order of the range of the interatomic forces is first
formed (owing to the interplay of the A-B interaction and the strain energy
accumulation) on top of which partially or completely relaxed 3D
islands nucleate and grow. The 3D islands and the thermodynamically stable
wetting layer represent necessarily different phases. If this were not the
case, the growth would continue by 2D layers. Thus we can consider as a useful
approximation to regard the 3D islanding on top of the uniformly strained 
wetting layer as a Volmer-Weber growth. 
That requires the mean adhesion of the atoms that 
belong to the base plane of the 3D islands to the stable wetting layer to be
smaller than the cohesion between them. In other words, the wetting of the
underlying wetting layer by the 3D islands must be incomplete. Otherwise, 3D
islanding will not occur.\cite{Rudy} In the Volmer-Weber growth the incomplete
wetting is due mainly to the difference in bonding ($E_{\rm AB} < E_{\rm AA}$),
the supplementary effect of the lattice misfit being usually smaller. In the
coherent SK growth ($E_{\rm AB} \approx E_{\rm AA}$), the incomplete wetting
is due to the lattice misfit, which leads to the atoms 
at the edges of the islands to displace from the bottoms of the potential
troughs provided by the atoms in the layer underneath.\cite{Merwe,Prieto}

As has been shown elsewhere,\cite{Kor,Prieto} it is the incomplete wetting 
which
determines the formation of dislocation-free 3D islands on top of the wetting
layer in the case of coherent SK growth. In this case, however, the relation
between $\Delta \mu $ and $\Phi $ is not as simple as given above. The bond
energies are generally not additive, the misfit strain is relaxed mostly 
near the side
and top walls and increasing the island's thickness leads to larger
displacements of the edge atoms from the bottoms of the potential troughs
provided by the wetting layer and in turn to a decrease of the wetting. 
For this reason, in this work we define the wetting parameter 
$\Phi $ as the difference of the interaction energies with the 
wetting layer of misfitting and non-misfitting 3D islands.

In the present paper we study the behavior of $\Phi $ for islands which 
belong to an array of islands. We study the effect of the density of the array
(the distance to the nearest neighbor islands), the size distribution (the
difference in size of the neighboring islands), and the shape distribution
(the slope of the side walls of the neighboring islands) on the wetting
parameter $\Phi $ of the considered island.

\section{Model}

\begin{figure}[htb]
\centering{\epsfysize=2cm\epsffile{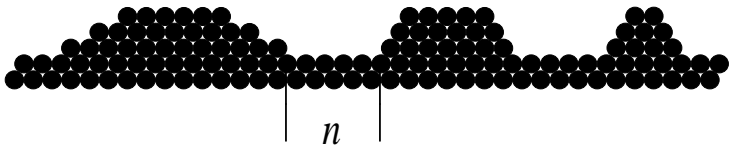}}
\caption{\label{array}
Schematic view of an array of islands on a wetting layer. The central island
is surrounded by two islands with different shapes and sizes. By $n$
is denoted the spacing between neighboring islands, which is a measure of the
density of the array.}
\end{figure}

We consider an atomistic model in $1+1$ dimensions (lateral size + height) 
which
we treat as a cross section of the real $2+1$ dimensional case. 
An implicit assumption is that in the real $2+1$ dimensional model 
the monolayer islands have a compact rather than a fractal shape and that 
the lattice misfit is the same in both orthogonal
directions. The 3D islands are represented by linear chains of atoms stacked
one upon the other.\cite{Stoop,Ratsch} Each upper chain is shorter than the
lower one. The shape of the islands in our model is given by the slope of the
side walls. For, example, an island
which consists of consecutive chains with N, N-1, N-2... atoms has a 60$^\circ$
slope of the side walls, whereas an island with chains consisting of, say, N,
N-5, N-10... atoms has a slope of 19.1$^\circ$, etc., where N is the number of
atoms in the base chain. The array in the 1+1 dimensional space is 
represented by a row of 3 or 5 islands on a wetting layer consisting of
several monolayers (Fig.~\ref{array}). The distance between two neighboring
islands is given by the number $n$ of vacant atomic positions betweeen the
ends of their base chains and can be varied from one to infinity. 

In order to simplify the computational procedure the ``wetting layer" in our
model is in fact composed of several monolayers of the true wetting layer
consisting of atoms of the overlayer material A plus several monolayers of
the unlike substrate material B. This composite wetting layer has the atom
spacing of the substrate material B as in the real case, but for the sake of
simplicity the atom bonding is that of the overlayer material A. 
We believe that
the latter does not introduce a perceptible error as the energetic influence
of the substrate B is screened by the true wetting layer A. We expect that
this approximation underestimates to some extent the wetting parameter $\Phi $
by making the composite substrate a bit softer than the real one (the A-A 
bonding is weaker than the B-B bonding,\cite{Merw0} for a later review
see Ref. \cite{Mar2}). We found that beyond 10 monolayers the studied
parameter $\Phi $ saturates its value. This is why in all cases given below,
unless otherwise stated, we allowed 10 monolayers to relax.

As in our previous work (Ref. \cite{Marjosem}), we make use of a simple
minimization procedure. The atoms interact through a potential that
can be easily generalized to vary its anharmonicity by adjusting
two constants $\mu $ and $\nu $ ($\mu  > \nu $) that govern separately the
repulsive and attractive branches, respectively,\cite{Mar3}
\begin{eqnarray}\label{potent}
V(x) = V_{o}\Biggl[\frac{\nu }{\mu - \nu }e^{-\mu (x-b)} - \frac{\mu }{\mu - 
\nu }e^{-\nu (x-b)}\Biggr],
\end{eqnarray}
where $b$ is the equilibrium atom separation. In this work, we have used
$\mu  = 2\nu $ with $\nu = 6 $, which turns the potential (\ref{potent}) 
into the familiar Morse form.

Our program calculates the interaction energy of all the atoms as well as its
gradient with respect to the atomic coordinates, i.e. the forces. 
Atoms in the islands and in the wetting layer are then allowed to relax.
Relaxation
of the system is performed by allowing the atoms to displace in the direction
of the gradient in an iterative procedure until the forces fall below some
negligible cutoff value. Periodic boundary conditions are applied in the
lateral direction.  We consider only interactions in the first coordination
sphere in order to mimic the directional bonds that are characteristic for
most semiconductor materials.\cite{Jerry}

\section{Results}

\begin{figure}[htb]
\centering{\epsfysize=5cm \epsffile{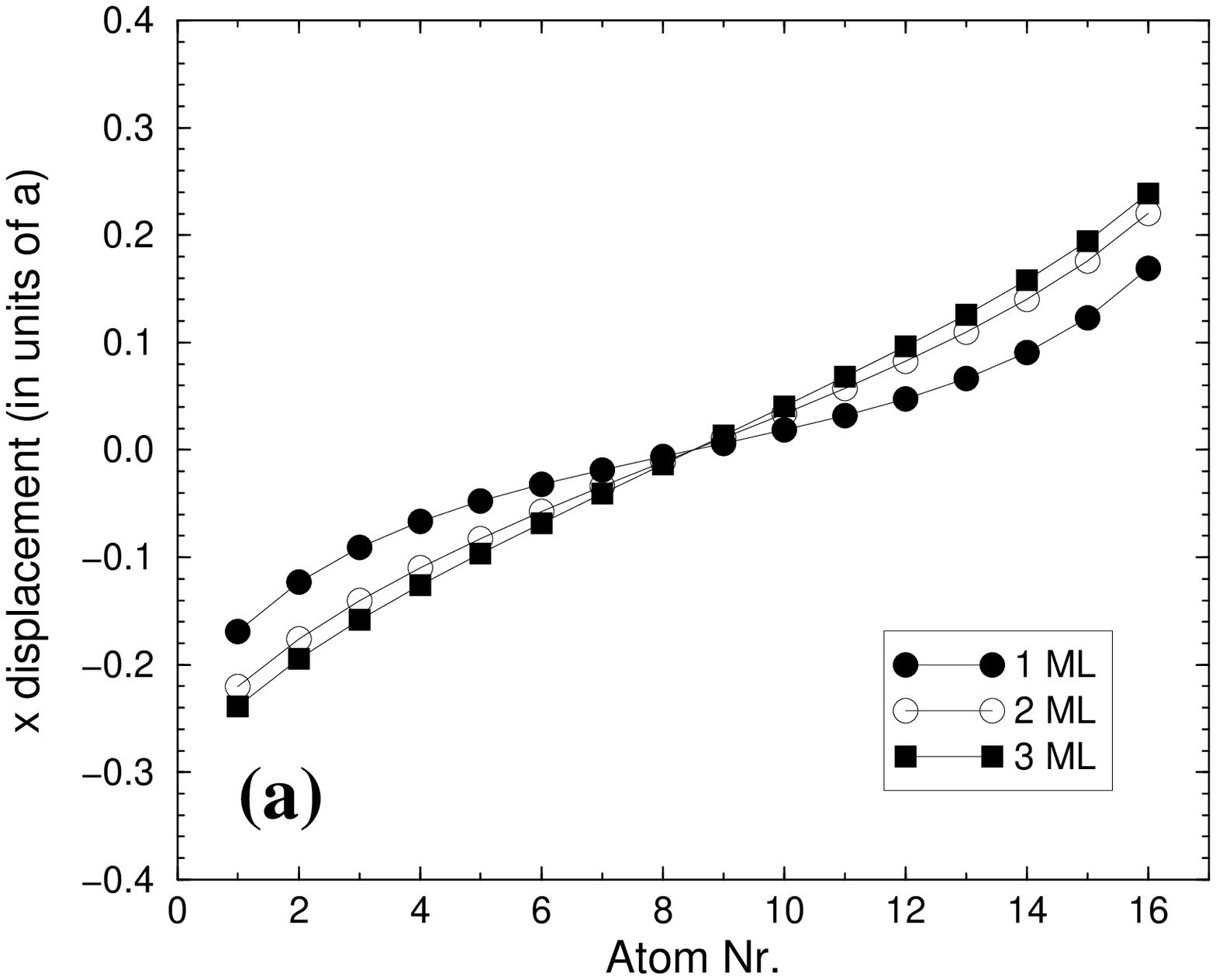}\epsfysize=5cm\epsffile{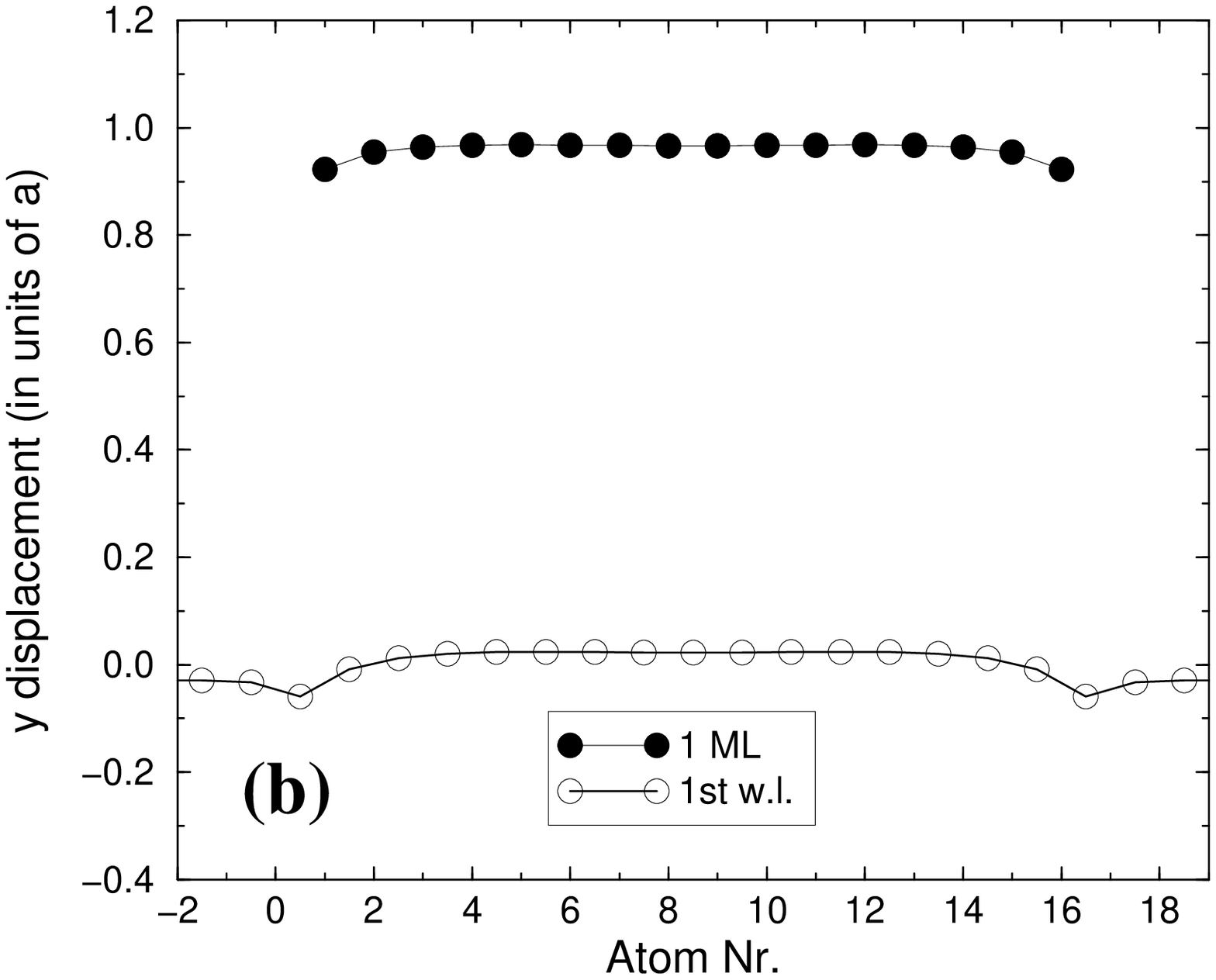} }
\caption{\label{displac}
Horizontal (a) and vertical (b) displacements of the atoms of the base chain
from the bottoms of the potential troughs provided by the homogeneously
strained wetting layer, for a lattice misfit of 7\%. The considered island
has 16 atoms in the base chain and is located between two identical ones
at a distance of $n = 5$. The displacements are given in
units of the lattice parameter $a$ of the composite wetting layer. The
horizontal displacements increase with increasing island thickness taken
in number of monolayers. Contrary to the rigid substrate case, the vertical
displacements of the edge atoms of the base chain of the islands and those of
the underlying atoms of the uppermost monolayer of the wetting layer, included
in (b), are directed downwards.}
\end{figure}

Figure~\ref{displac}(a) shows the horizontal
displacements of the atoms of the base chain from the bottoms of the potential
troughs provided by the homogeneously strained wetting layer for a misfit of
7\%. The considered island has two identical ones at a distance of
$n = 5$. This is the same behavior as predicted by the one-dimensional model of
Frank and van der Merwe.\cite{Merwe} The horizontal displacements increase
with increasing island thickness (measured in number of monolayers) precisely
as in the case of a rigid substrate and non-interacting islands.\cite{Marjosem}
In contrast to the rigid substrate case,\cite{Prieto} the vertical 
displacements of the edge atoms of the base chain of the islands 
and the underlying atoms of the uppermost monolayer of the
wetting layer are directed downwards (Fig.~\ref{displac}(b)). It is worth
noting that the same result has been found by Lysenko {\it et al.} in the
case of homoepitaxial metal growth by using a computational method within the
framework of the tight-binding model.\cite{Lys}

\begin{figure}[htb]
\centering{\epsfysize=6cm \epsffile{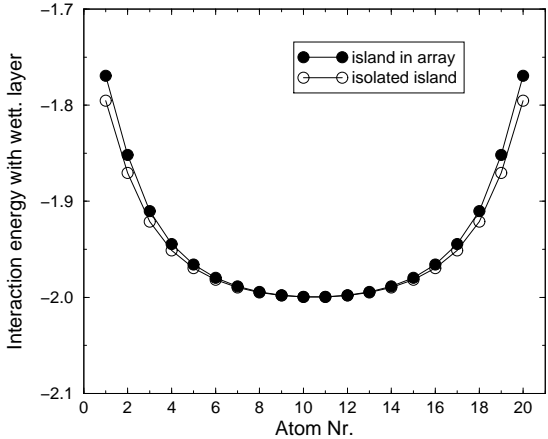}}
\caption{\label{vsi}
Distribution of the interaction energy (in units of $V_0$)
between the atoms of the base chain A of
a 3~ML-high, coherent island with 20 atoms in the base chain, and the
underlying wetting layer B, for a positive misfit of 7\%. Full circles
correspond to an island separated by a distance $n = 5$ from two identical
ones, the empty ones correspond to a reference isolated island.}
\end{figure}

In spite of their downwards vertical displacements, the edge atoms are again
more weakly bound to the underlying wetting layer, as in the rigid
substrate models of Refs. \cite{Kor} and \cite{Marjosem} (Fig.~\ref{vsi}). 
Shown in the same
figure for comparison is also an island without neighbors. As seen, the edge
atoms of the single island adhere more strongly to the substrate.  
This is in fact the essential physics behind the effect 
that the neighboring islands exert on the middle island. 
The central island looses to some degree contact with the substrate 
(in this case the wetting layer) and the wetting parameter is increased. 
We can interpret this as the wetting layer becoming stiffer under the 
influence of the neighboring islands.

\begin{figure}[htb]
\centering{\epsfysize=6cm \epsffile{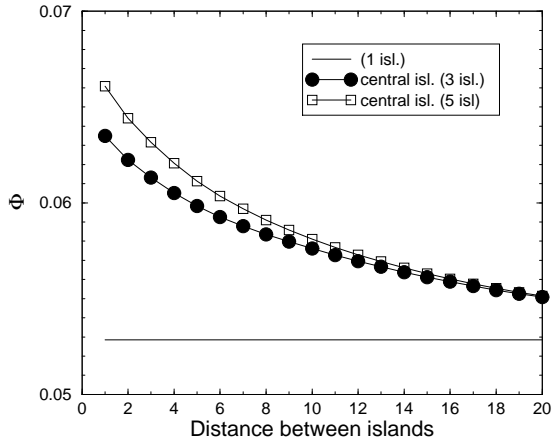}}
\caption{\label{phivsn}
Dependence of the wetting parameter of the central island on the distance $n$
between the islands. Results for arrays of 3 and 5 islands are given, as
well as for a reference isolated island. All islands are 3~ML-high, have
20 atoms in their base chains and the lattice misfit amounts to 7\%.
As seen, the next-nearest neighbors
play a smaller but not negligible role.}
\end{figure}

The influence of the density of the array is demonstrated in Fig.~\ref{phivsn}. The values for 3 and 5 islands were calculated assuming
equally spaced islands. These can be thus treated as a self-organized array.
As expected the wetting parameter increases with decreasing
distance between the islands or, in other words, with increasing
array density. 

\begin{figure}[htb]
\centering{\epsfysize=6cm \epsffile{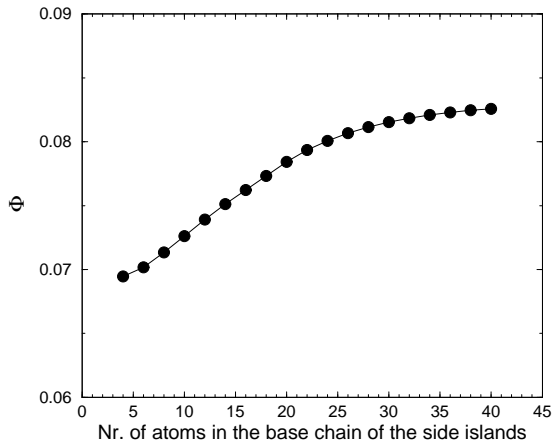}}
\caption{\label{phivsv}
Dependence of the wetting parameter of the central island on the size
of the base chains of the two side islands. These two have the same
volume and are separated from the central one by a distance $n$ = 5.
All the islands are 3 ML high, the central one having 20 atoms in the
base chain, the misfit amounts to 7\% and the wetting layer consists of
3~ML which are allowed to relax.}
\end{figure}

\begin{figure}[htb]
\centering{\epsfysize=6cm \epsffile{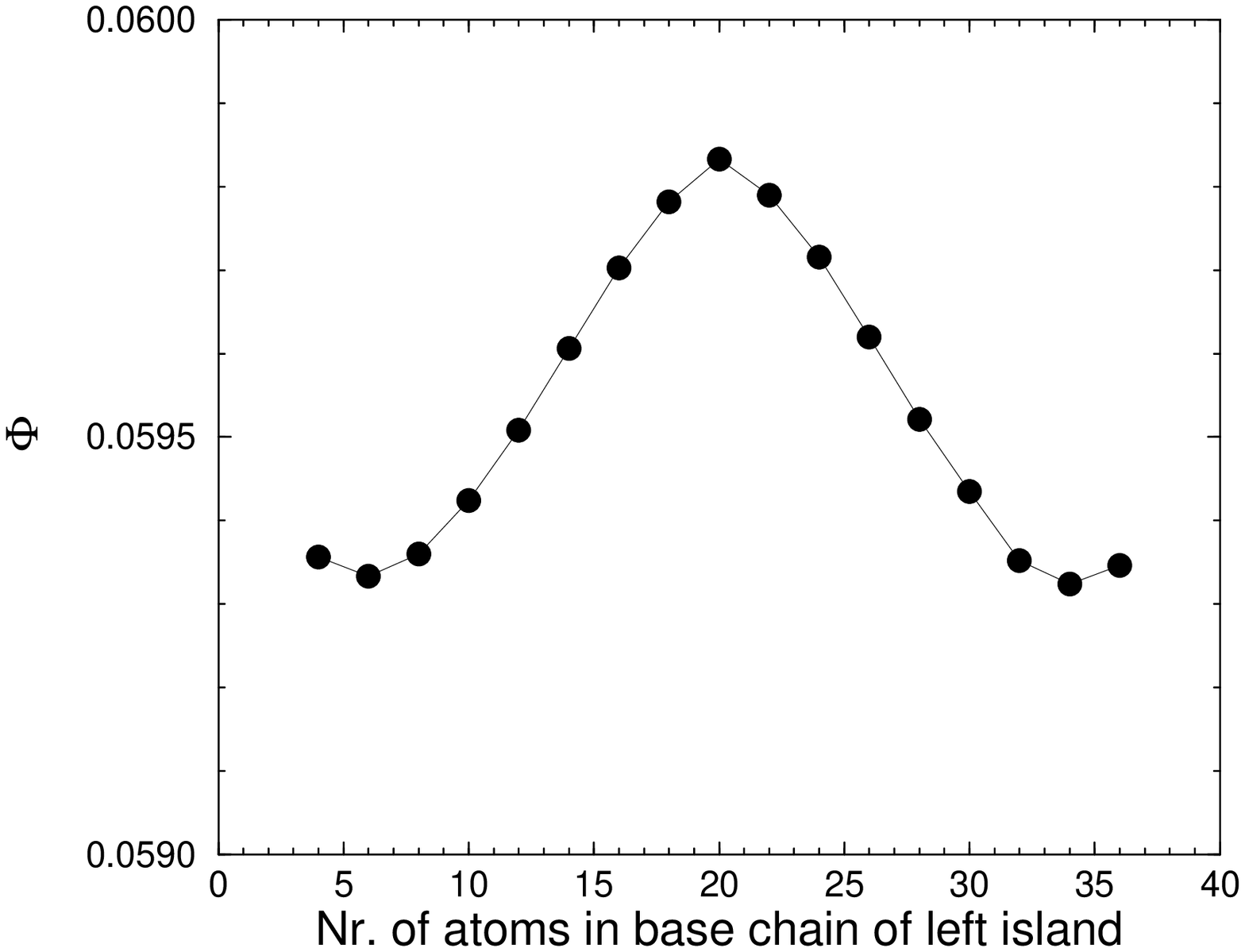}}
\caption{\label{phyasym}
Dependence of the wetting parameter of the central island on the size
distribution of the side islands. The central island is three monolayers
thick and has 20 atoms in the base chain thus containing a total of 57 atoms.
The lattice misfit is 7\%.
The x-axis represents the number of atoms in the base chain of the left
island. The sum of the volumes of the left and
right islands is kept constant and equal to the doubled volume
of the central island. We increase the volume of the left island and
decrease the volume of the right island thus passing through the middle
point at the maximum where the three islands have equal volumes.}
\end{figure}

Figure~\ref{phivsv} shows the wetting parameter of the central island as
a function of the size of the side islands. For this calculation, we
considered three islands with the same thickness of 3 monolayers.
Furthermore, both side islands have one and the same volume. Increasing
the volume of the side islands leads to an increase of the elastic fields
around them and to a further reduction of the bonding between the edge 
atoms of the central island and the wetting layer.

\begin{figure}[htb]
\centering{\epsfysize=6cm \epsffile{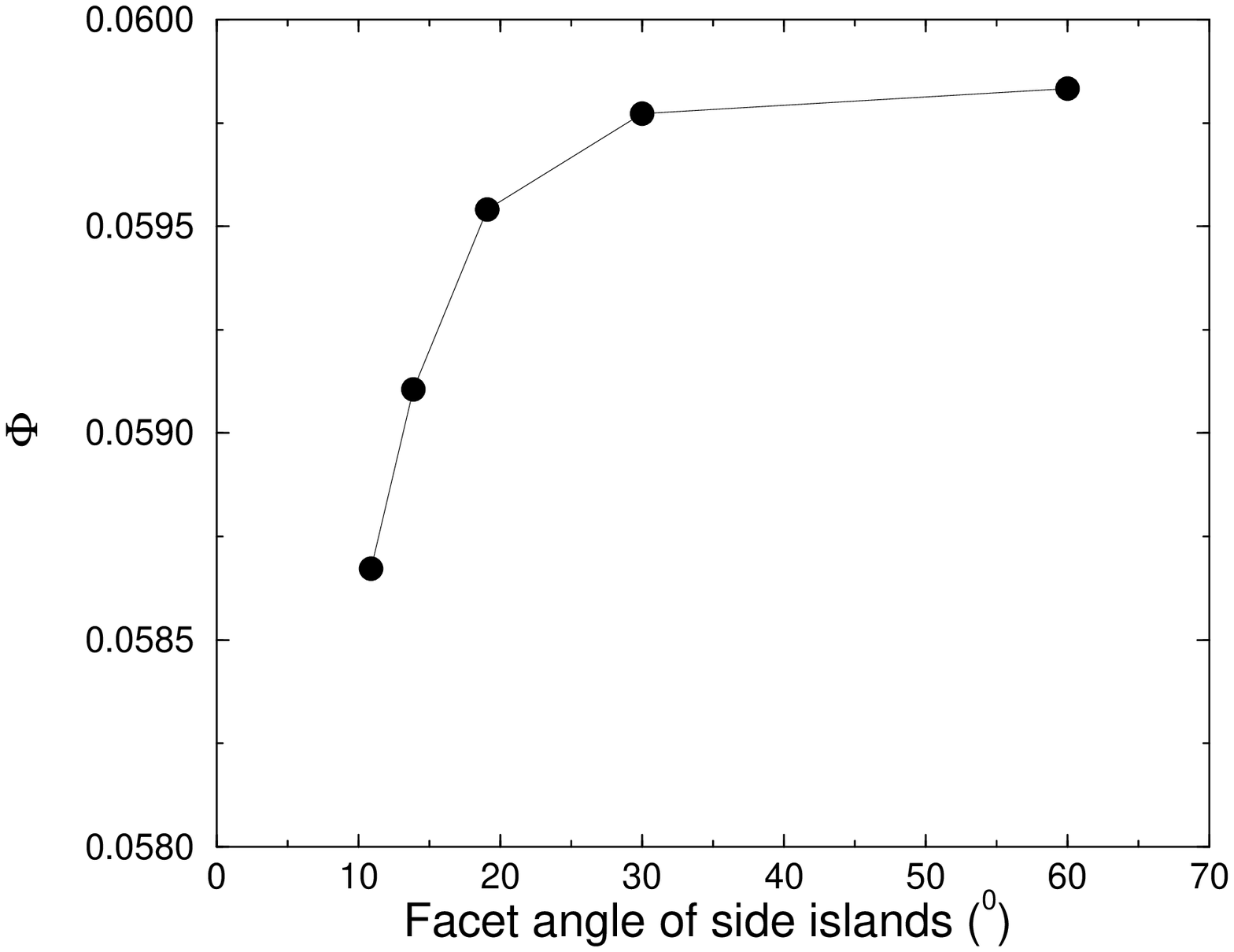}}
\caption{\label{phi60d}
Dependence of the wetting parameter of the central island on the shape of the
neighboring islands, measured in degrees of their facet angles. The central
island has a slope of 60$^\circ$ of its side walls. All islands are 3 ML high,
have 20 atoms in their base chains and are separated by a distance $ n = 5 $
The lattice misfit amounts to 7\%.}
\end{figure}

Figure~\ref{phyasym} demonstrates one of the most important results, the
effect of the size distribution on the wetting of the islands. It shows the
behavior of the wetting parameter $\Phi $ of the central island as a function
of the number of atoms in the base chain (which is a measure of the volume) of
the left island. In this case, the sum of the volumes (the total number of 
atoms) of the left and right islands is kept constant and precisely equal 
to the doubled volume of the central island. 
All three islands have the same thickness.
The side walls of all the three islands make an
angle of 60$^\circ$ with their bases. Thus the first point (and, by symmetry,
also the last one) gives the maximum
asymmetry in the size distribution of the array, the left island consists of
9 atoms whereas the right island is built of 105 atoms. The point at the
maximum of wetting describes the monodisperse distribution -- the three
islands have one and the same volume of 57 atoms.  
As seen in the case of perfect self-assembly of the 
array the wetting parameter, or in other words, the tendency to clustering 
displays a maximum value.

The effect of the shape of the side islands, i.e. their facet angles, on the
wetting parameter of the central island is demonstrated in Fig.~\ref{phi60d}.
The slope of the facets of the central island is 60$^\circ$. The effect is
greatest when the side islands have the steepest walls. The same result (not
shown) is obtained when the central island has a different facet angle,
e.g. 11$^\circ$.
The explanation follows the same line as the one given above. 
The side islands with larger-angle side walls exert a greater elastic 
effect on the wetting layer and in turn on the displacements and the bonding 
of the edge atoms of the central island. 

\begin{figure}[htb]
\centering{\epsfysize=6cm \epsffile{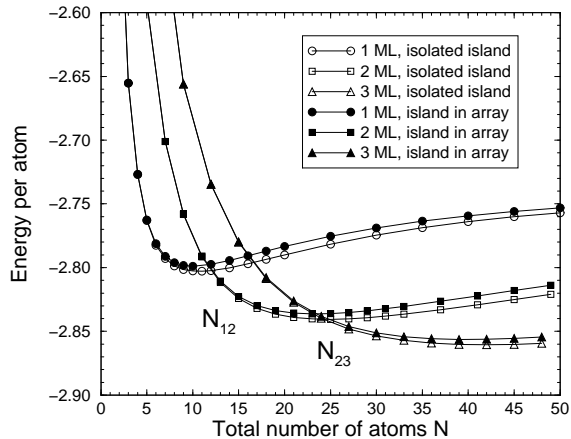}}
\caption{\label{evsn123}
Dependence of the energy per atom on the total number of atoms in
compressed coherently strained islands with different thicknesses in
monolayers denoted by the figures at each curve for a misfit of 7\%.
The considered island has two identical neighbours at a distance $n = 5$.
The analogous curves for single isolated islands (empty symbols) are also
given for comparison. The wetting layer consists in all cases of 3~ML
which are allowed to relax.}
\end{figure}

We studied also the stability of islands with a thickness increasing by one
monolayer in the presence of two islands on both sides on a deformable
substrate. 
The result in Fig.~\ref{evsn123} shows the same behavior observed
in Refs. \cite{Kor} and \cite{Marjosem} where rigid substrates were assumed.
This means that the overall transformation from the precursor 2D islands
to a macroscopic 3D islands takes place in consecutive stages in each of which
the islands thicken by one monolayer. As shown in Refs. \cite{Kor} and
\cite{Marjosem} the latter leads in turn to a critical misfit beyond which
coherent 3D islanding takes place, and below which the lattice mismatch is
accommodated by misfit dislocations. The existence of a critical misfit has
been experimentally observed in a series of different
systems.\cite{Leo,Xie,Wal,Pinc} The energies computed in the case of the
reference single islands always lie below the curves of the islands in an array.
The difference obviously gives the energy of repulsion between the neighboring
islands. It follows from the above that the presence of neighboring islands
leads to a slight decrease of the critical misfit.

\section{Discussion}

For the discussion of the above results we have to bear in mind that 
a positive wetting parameter shows in fact a tendency of the deposit 
to form 3D clusters rather than a planar film. 
In the case of coherent SK growth, the non-zero
wetting parameter is due to the weaker adhesion of the atoms that are closer
to the islands edges. The presence of other islands, particularly with large
angle facets, in the near vicinity of the considered island 
makes this effect stronger
as seen in Fig.~\ref{vsi}. The transformation of two-dimensional islands
with a monolayer height into bilayer three-dimensional islands takes place by
detachment of atoms from the edges and their subsequent jumping and collision
on the top island's surface.\cite{Stmar} Thus this edge effect clearly
demonstrates the influence of the lattice misfit on the rate of second layer
nucleation and in turn on the kinetics of the 2D-3D
transformation.\cite{Fil,Lin} The presence of neighboring islands facilitates
and thus accelerates the formation of 3D clusters and their further growth. In
a self-assembled population of islands the tendency to clustering
is thus enhanced.

We can think of the flatter islands in our model (11$^\circ$ facet angle) as
the famous ``hut'' clusters discovered by Mo {\it et al.},\cite{Mo} and of the
clusters with 60$^\circ$ facet angles as the ``dome'' clusters. It is well known
that clusters with steeper side walls relieve the strain much more efficiently
than the flatter clusters (see the discussion in Ref. \cite{Mich}; the planar
film, which is the limiting case of the flatter islands with a facet angle 
equal to zero, does not relieve the strain at all.) 
We see that large-angle facet
islands affect more strongly the growth of the neighboring
islands, leading thus to a more narrow size distribution.

We further conclude that a self-assembled population of quantum dots is
expected at comparatively low temperatures such that the critical 
wetting-layer thickness for 3D islanding to take place approaches 
an integer number
of monolayers. In InAs/GaAs quantum dots, the reported values of the critical
thickness were found to vary from 1.2 to 2 monolayers,\cite{Pol} (see also
the discussion in Ref. \cite{Scheff} and the references therein). The
critical wetting-layer thickness should be given by an integer number of
monolayers plus the product of the 2D island density and the critical volume
(or area) $N_{12}$. The 2D island
density increases steeply with decreasing temperature.\cite{dobbs} In such a
case, a dense population of 2D islands will overcome simultaneously the
critical size $N_{12}$ to produce 3D bilayer islands. The latter will
interact maximally with each other from the very beginning 
of the 2D-3D transformation giving rise to a maximum wetting parameter 
and, in turn, to large-angle facets and a narrow size distribution. 
This is in agreement with the observations of
Le Tanh {\it et al.} in the case of nucleation and growth of self-assembled
Ge quantum dots on Si(001).\cite{Le} At 700$^\circ$C, a population of islands
with a concentration of the order of 1$\times 10^7$ - 1$\times 10^8$ cm$^{-2}$
is obtained; the islands have the shape of a truncated square pyramid with four
side wall facets formed by (105) planes with an inclination angle of
about 11$^\circ$ and the size distribution of the islands is quite broad. 
On the other hand, at
550$^\circ$C, a population of islands with an areal density of the order
of 1$\times 10^9$ - 1$\times 10^{10}$ cm$^{-2}$ is observed, the
islands have larger angle (113) facets and their size distribution 
is much more narrow. 

In summary, we have shown that the presence of neighboring islands 
decreases the wetting of the substrate (in this case the wetting layer) 
by the 3D islands. The larger the density of
the array, the weaker the wetting. Neighboring islands with steeper side
walls reduce more strongly the wetting of the considered island. The wetting
parameter displays a maximum (implying a minimal wetting) when the array 
shows a 
monodisperse size distribution. We should expect optimum self-assembled 
islanding
at lower temperatures such that the 2D-3D transformation takes place at
the maximum possible island density.

\acknowledgements
J.E.P. gratefully acknowledges financial support from the 
Alexander-von-Humboldt Stiftung and the Spanish MEC
(grant No. EX2001 11808094).

\end{document}